\documentstyle[preprint,aps,epsf]{revtex}
\begin{document}

\tightenlines

\def\lqcd{\Lambda_{\rm QCD}}
\def\xslash#1{{\rlap{$#1$}/}}
\def\dsl{\,\raise.15ex\hbox{/}\mkern-13.5mu D}
\preprint{\vbox{\hbox{UTPT-- 99-11}
\hbox{hep-ph/9907517}}}

\def\ctp#1#2#3{\CTP{\bf #1} (#2) #3}
\def\jetpl#1#2#3{\JETPL{\bf #1} (#2) #3}
\def\nc#1#2#3{\NC{\bf #1} (#2) #3}
\def\np#1#2#3{\NP{\bf B#1} (#2) #3}
\def\pl#1#2#3{\PL B {\bf #1} (#2) #3}
\def\prl#1#2#3{\PRL{\bf #1} (#2) #3}
\def\prd#1#2#3{\PR D {\bf #1} (#2) #3}
\def\prep#1#2#3{\PRep{\bf #1} (#2) #3}
\def\physrev#1#2#3{\PR{\bf #1} (#2) #3}
\def\sjnp#1#2#3{\SJNP{\bf #1} (#2) #3}
\def\nuvc#1#2#3{\NC{\bf #1A} (#2) #3}
\def\blankref#1#2#3{   {\bf #1} (#2) #3}
\def\ibid#1#2#3{{\it ibid,\/}  {\bf #1} (#2) #3}
\def\AP{{\it Ann.\ Phys.\ }}
\def\CMP{{\it Comm.\ Math.\ Phys.\ }}
\def\CTP{{\it Comm.\ Theor.\ Phys.\ }}
\def\IJMP{{\it Int.\ Jour.\ Mod.\ Phys.\ }}
\def\JETPL{{JETP Lett.\ }}
\def\NC{{\it Nuovo Cimento\ }}
\def\NP{{Nucl.\ Phys.\ }}
\def\PL{{Phys.\ Lett.\ }}
\def\PR{{Phys.\ Rev.\ }}
\def\PRep{{Phys.\ Rep.\ }}
\def\PRL{{Phys.\ Rev.\ Lett.\ }}

\title{Nonperturbative corrections to $B \to X_s \ell^+ \ell^-$ with 
phase space restrictions}

\author{Christian W. Bauer and Craig N. Burrell}

\address{
\medskip Department of Physics, University of Toronto\\60
  St.~George Street, Toronto, Ontario,
  Canada M5S 1A7 
\medskip}

\bigskip
\date{July 1999} 

\maketitle

\begin{abstract} We study nonperturbative corrections up to ${\cal
  O}(1/m_b^3)$ in the inclusive rare $B$ decay $B \to X_s \ell^+ \ell^-$
  by performing an
  operator product expansion. The values of the matrix elements 
  entering at this order are unknown and introduce uncertainties into
  physical quantities.
  Imposing a phase space cut to
  eliminate the $c \bar{c}$ resonances we find that the  ${\cal
  O}(1/m_b^3)$ corrections introduce an ${\cal O}(10\%)$ uncertainty
  in the measured rate. We also find that the contributions arising at 
  ${\cal O}(1/m_b^3)$ are comparable to the ones arising at ${\cal
  O}(1/m_b^2)$ over the entire region of phase space.
\end{abstract}

\vskip2cm

By 1995 CLEO had measured the rates for both the exclusive decay $B
\to K^* \gamma$ \cite{CLEOex} and for the inclusive process $B \to X_s
\gamma$ \cite{CLEOin}, marking the advent of experimental studies of
penguin--mediated 
$B$ decays.  Such processes arise in the standard model at 
the one loop level. Physics from beyond the standard model may appear in 
the loop with an
amplitude comparable to the standard model amplitude, thereby making
such rare decays an excellent testing ground for standard model extensions. 
Of course, efforts to detect deviations from 
the standard model are frustrated by uncertainties in standard model 
predictions. 

The decay $ B \rightarrow X_s \ell^+ \ell^- $,
though it has not yet been observed \cite{CLEO_search}, has
garnered recent interest because of its sensitivity to new physics
not contributing to the decay $B \to X_s \gamma$.  The ${\cal O}(1/m_b^2)$ 
nonperturbative corrections to $\Gamma( B \rightarrow X_s
\ell^+ \ell^- )$ have been previously 
calculated \cite{hiller_hqet,hiller_hadron}. 
We extend that study to calculate the
${\cal O}(1/m_b^3)$ corrections for massless leptons in the final
state $(\ell = e,\mu)$ following the similar calculations for semileptonic $B$ 
decays \cite{gremm} and the decay $B \to X_s \gamma$
\cite{christian}. We use the standard effective 
Hamiltonian mediating the
$b(p_b) \to s(p_s) + \ell^+(p_+) + \ell^-(p_-)$
transition obtained from integrating out the top quark and the weak
bosons. It is given by
\begin{equation}
{\cal H}_{eff}(b \to s \ell^+ \ell^-) = -4 \frac{G_F}{\sqrt{2}}
 |V_{ts}^* V_{tb}| \sum_{i=1}^{10} C_i(\mu) O_i(\mu).
\end{equation}
The operator basis $\{O_i\}$ can be found in the literature
\cite{b->see}.

The Wilson coefficients $\{ C_i \}$ at the scale $\mu \sim m_b$ are
known in the next to leading log approximation \cite{misiak,buras-munz}.
For consistency with
the literature we have defined two effective Wilson coefficients: 
$C_7^{eff} \equiv C_7 - C_5/3 - C_6$ and $C_9^{eff}$. The latter contains 
the operator mixing of $O_{1-6}$ into $O_9$ as well as the one loop
matrix elements of $O_{1-6,9}$ \cite{misiak,buras-munz}.  The 
full analytic expression for $C_9^{eff}$ is quite
lengthy and may be found in \cite{buras-munz}.

For the branching ratio at the parton level we find, in agreement 
with previous calculations \cite{hiller_hadron,b->see},
\begin{eqnarray}
\frac{{\cal B}_{\rm parton}}{{\cal B}_0} =  - \frac{32}{9} \left( 4 + 3
    \log\left( \frac{4 m_l^2}{M_B^2} \right) 
  \right) C_7^{eff^2} &+& \frac{2}{3} C_{10}^2 + 
  128\, C_7^{eff} \int_0^\frac{1}{2} dx_0 
   \; x_0^2 \; C_9^{eff}(x_0)  \nonumber \\
  &+& \frac{32}{3} \int_0^\frac{1}{2} dx_0 \left( 
    3 x_0^2 -4 x_0^3 \right) | C_9^{eff}(x_0) |^2
\label{parton_br}
\end{eqnarray}
where $x_0 \equiv E_0/m_b$ is the rescaled final state parton energy, 
and ${\cal B}_0$ is the normalization factor
\begin{equation}
{\cal B}_0 = {\cal B}_{sl} \frac{3\alpha^2}{16 \pi^2} \frac{\left|
    V_{ts}^*V_{tb}\right|^2}{\left|V_{cb} \right|^2}
\frac{1}{f(\hat{m}_c)\kappa(\hat{m}_c)}.
\end{equation}
Here ${\cal B}_{sl}$ is the measured semileptonic branching ratio,
$f(\hat{m}_c)$ is the phase space factor for $\Gamma(B \to X_c \ell \bar{\nu})$
\begin{equation}
f(\hat{m}_c) = 1 - 8 \hat{m}_c^2 + 8 \hat{m}_c^6 - \hat{m}_c^8 - 24
\hat{m}_c^4 \log(\hat{m}_c),
\end{equation}
and $\kappa(\hat{m}_c)$ accounts for the ${\cal O}(\alpha_s)$ QCD
correction and the leading power corrections. The complete expression
for $\kappa(\hat{m}_c)$ may be found in \cite{hiller_hqet}.
As alluded to above, the analytic form of $C_9^{eff}$ is sufficiently 
complicated that we must resort to numerical integrations.     

The procedure for calculating nonperturbative contributions to
heavy hadron decays has been discussed in great detail in the
literature \cite{chay,inclusive}, and we give only a short review
here. The differential rate is proportional to the product of the lepton tensor
$L_{\mu\nu}$ and the hadron tensor $W^{\mu\nu}$, which for the process
in question may be written as
\begin{equation}
d\Gamma = \frac{1}{2M_B} \frac{G_F^2 \alpha^2}{2\pi^2} |V_{ts}^* V_{tb}|^2 
d\Pi \left( L_{\mu\nu}^L W^{L\mu\nu} + L_{\mu\nu}^R W^{R\mu\nu} \right)
\end{equation}
where $\Pi$ denotes the three body phase space.
The hadron tensor $W^{\mu\nu}$ is related via the optical theorem to the 
imaginary part of the forward scattering matrix element 
$W^{\mu\nu} = 2 \,{\rm Im} T^{\mu\nu}$ where  
\begin{equation} 
\label{discontinuity}
T^{L(R)}_{\mu\nu} = - i \int d^4 x \, e^{-i q \cdot x} 
  \left\langle B \left| T\{ J^{L(R)^\dagger}_\mu(x) ,J^{L(R)}_\nu(0)\}
  \right| B \right\rangle
\end{equation}
and the spin-summed tensor $L_{\mu\nu}$ for massless leptons is
\begin{equation}
L^{L(R)}_{\mu\nu} = 2 \left[ p_+^\mu p_-^\nu + p_-^\mu  p_+^\nu  - 
g^{\mu\nu} p_+\cdot p_- \mp i \epsilon^{\mu\nu\alpha\beta}p_{+\alpha}
p_{-\beta} \right].
\end{equation}
In (\ref{discontinuity}) $J^\mu$ denotes the current mediating this
transition, and is given by
\begin{equation}
J^\mu_{L(R)} = \bar{s} \left[ R \gamma^\mu \left( C_9^{eff} \mp C_{10} + 
  2 C_7^{eff} \frac{ \rlap /\hat{q} }{\hat{q}^2} \right) + 
  2 \hat{m}_s C_7^{eff} \gamma^\mu \frac{ \rlap /\hat{q} }{\hat{q}^2} L
  \right] b
\end{equation}
where $L(R) = \frac{1}{2} (1 \mp \gamma_5)$ are the usual left and
right handed chiral 
projection operators, and $q \equiv (p_+ + p_-)$ is the dilepton momentum.
 
It has been shown in \cite{chay,inclusive} that the time--ordered product in 
(\ref{discontinuity}) can be expanded as an operator product expansion 
(OPE), given schematically by
\begin{equation}
- i \int d^4 x \, e^{-i q \cdot x} 
  T\{ J^\dagger(x) ,J(0)\} \sim \frac{1}{m_b} \left[ O_0 + \frac{1}{2m_b} O_1 +
  \frac{1}{4m_b^2} O_2 + \frac{1}{8m_b^3} O_3 + \ldots \right],
\label{ope}
\end{equation}
where $O_n$ represents a set of local operators of dimension $(3+n)$. In
this study we include operators up to dimension six \cite{christian}. 

Matrix elements of dimension four operators
vanish \cite{chay} at leading order in the $1/m_b$ expansion and
matrix elements of dimension five operators may be
parameterized by $\lambda_1$ and $\lambda_2$ \cite{paramd2} 
\begin{equation}
  \langle B(v) | \bar{h}_v \Gamma iD_\mu iD_\nu h_v | B(v) \rangle = M_B {\rm
  Tr} \left\{\Gamma P_+ \left(\frac{1}{3} \lambda_1 (g_{\mu\nu} - v_\mu v_\nu)
  +\frac{1}{2} \lambda_2 i \sigma_{\mu\nu}\right) P_+\right\},
\label{lambda_gen}
\end{equation}
where $P_+=\frac{1}{2}(1+\rlap/v)$ projects
onto the effective spinor $h_v$, and $\Gamma$ is an arbitrary Dirac 
structure. 

Finally, the dimension six operators may
be parameterized by the matrix elements of two local operators 
\cite{gremm,mannel} 
\begin{eqnarray}\label{rho}
\frac{1}{2M_B}\langle B(v)| \bar{h}_v iD_\alpha iD_\mu iD_\beta
h_v | B(v)\rangle&=&\frac{1}{3}\rho_1\left(g_{\alpha\beta}-v_\alpha
v_\beta\right) v_\mu, \nonumber\\
 \frac{1}{2M_B}\langle B(v)| \bar{h}_v
iD_\alpha iD_\mu iD_\beta \gamma_\delta \gamma_5 h_v
| B(v)\rangle&=&\frac{1}{2} \rho_2 \, i\epsilon_{\nu\alpha\beta\delta}
v^\nu v_\mu 
\end{eqnarray}
and by matrix elements of two time--ordered products
\begin{eqnarray}\label{tau}
\frac{1}{2 M_B} \langle B(v)|\bar{h}_v (iD)^2h_vi\int
d^3x\int_{-\infty}^0 \!\!\!\! dt \; {\cal
L}_I(x)| B(v)\rangle+h.c.&=&\frac{{\cal T}_1 + 3 {\cal
T}_2}{m_b}, \nonumber\\ 
\frac{1}{2M_B} \langle B(v)|\bar{h}_v
\frac{1}{2}(-i \sigma_{\mu\nu})G^{\mu\nu} h_vi\int d^3x\int_{-\infty}^0
  \!\!\!\! dt\; {\cal L}_I(x)| B(v)\rangle+h.c.&=&\frac{{\cal T}_3+3{\cal
T}_4}{m_b}.
\end{eqnarray}
The contributions from ${\cal T}_{1-4}$ can most easily be 
incorporated by making the replacements \cite{gremm}
\begin{eqnarray}
\lambda_1 &\to& \lambda_1 + \frac{{\cal T}_1 + 3 {\cal T}_2}{m_b}
\nonumber\\
\lambda_2 &\to& \lambda_2 + \frac{{\cal T}_3 + 3 {\cal T}_4}{3 m_b}
\label{tsubs}
\end{eqnarray}
in the final analytic results.
In addition, there is a contribution to the total rate from the four--fermion operator 
\begin{equation}
O_{(V-A)}^{bs} = 16 \pi^2 \; \left[ \bar{b} \gamma^\mu L s \bar{s} \gamma^\nu L b \;
                 ( g_{\mu\nu} - v_\mu v_\nu ) \right],
\end{equation}
the matrix element of which we define as
\begin{equation}
\frac{1}{2 M_B} \langle B | O_{(V-A)}^{bs} | B \rangle \equiv f_1.
\end{equation}

\begin{figure}
\centerline{\epsfxsize=13 cm \epsfbox{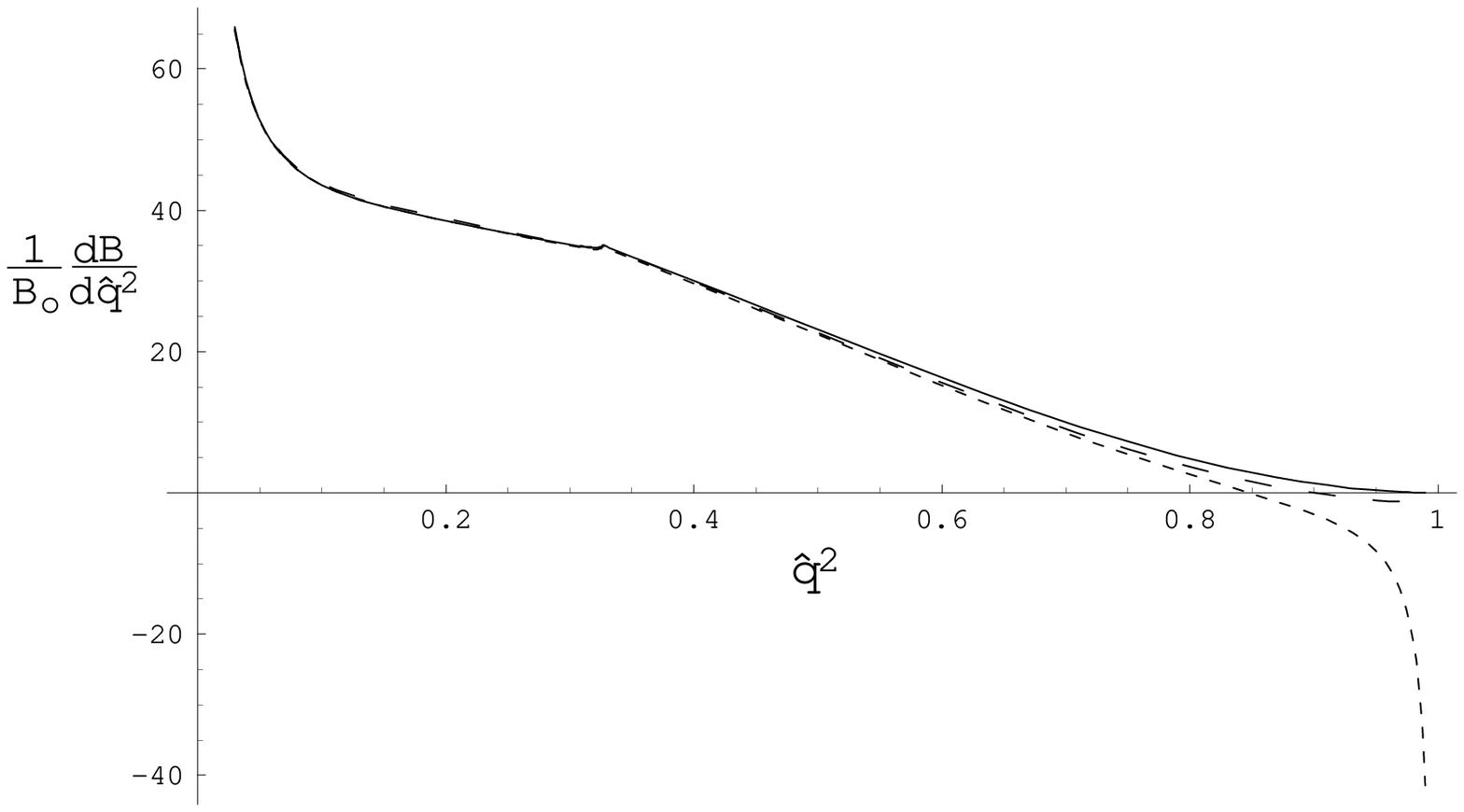}}
\caption{The differential decay spectrum. The solid line shows the
  parton model prediction, the dashed line includes the ${\cal O}(1/m_b^2)$
  corrections and the dotted line contains all corrections up to $
  {\cal O}(1/m_b^3)$.}
\label{diff_spec}
\end{figure}

Our analytic expression for the differential branching ratio agrees
with the results presented in \cite{hiller_hqet} up to 
${\cal O}(1/m_b^2)$, and will be presented in detail elsewhere \cite{bigone}. 
Here we restrict ourselves
to numerical results. In Figure \ref{diff_spec} we plot the differential 
branching ratio $\frac{1}{{\cal B}_0} \frac{d{\cal B}}{d\hat{q}^2}$, 
where $\hat{q} = q/m_b$. 
The solid line is the parton model
result, the long-dashed line includes corrections up to ${\cal O}(1/m_b^2)$ 
and the short-dashed line incorporates typical ${\cal O}(1/m_b^3)$ corrections 
as well. In
these plots we have used $\lambda_2 = 0.12 \,\,{\rm GeV}^2$ as indicated by the $B^*-B$ mass 
splitting, and $\lambda_1 = -0.19 \,\,{\rm GeV}^2$ \cite{lam1values}. For the
${\cal O}(1/m_b^3)$ matrix elements, whose values are unknown, we have 
chosen a generic size  
$|\rho_i|,\,|{\cal T}_i| \sim \lqcd^3 \sim (0.5 \,\,{\rm GeV})^3$
as suggested by dimensional analysis. 

 Compared to the parton model prediction the
nonperturbative corrections are small over almost the entire range of
$\hat{q}^2$, and become large only near the $\hat{q}^2 \to 1$
endpoint. It is a well known feature 
that close to this endpoint the
OPE (\ref{ope}) breaks down and the differential spectrum is determined by the 
shape function \cite{shape_func}. Once this spectrum is smeared with
a weight function that varies slowly in the endpoint region the OPE
should be convergent. However, as can be seen from Figure \ref{diff_spec} 
the differential branching ratio diverges in the $\hat{q}^2 \to 1$ endpoint
as 
\begin{equation}
\left. \frac{1}{{\cal B}_0} \frac{d {\cal B}}{d\hat{q}^2}
\right|_{\hat{q}^2 \to 1} \sim \frac{\rho_1}{1 - \hat{q}^2},
\end{equation}
yielding an unphysical logarithmic divergence in the integrated
spectrum 
that is 
regulated by the $s$ quark mass. This apparent problem is solved by
an additional term that contributes only at the endpoint
\begin{equation}
\frac{d\Gamma}{d\hat{q}^2} = \frac{d\Gamma}{d\hat{q}^2}|_{reg} - 
 16 \; (C_{10}^2 + (2 \, C_7^{eff} + C_9^{eff})^2) \,\,\delta ( q^2 -
 1) \left( \rho_1 \log(\hat{m}_s) - f_1
  \right),
\end{equation}
where $\frac{d\Gamma}{d\hat{q}^2}|_{reg}$ is the function plotted in
Figure \ref{diff_spec}.  The $\log(\hat{m}_s)$ term multiplying the delta function 
removes the divergence
mentioned above and the appearance of the four fermion operator has
been discussed in the context of semileptonic $B$ decays
\cite{sumlogs,blok}. A more detailed discussion of these issues will
be given elsewhere \cite{bigone}.

\begin{figure}
\centerline{\epsfxsize=13 cm \epsfbox{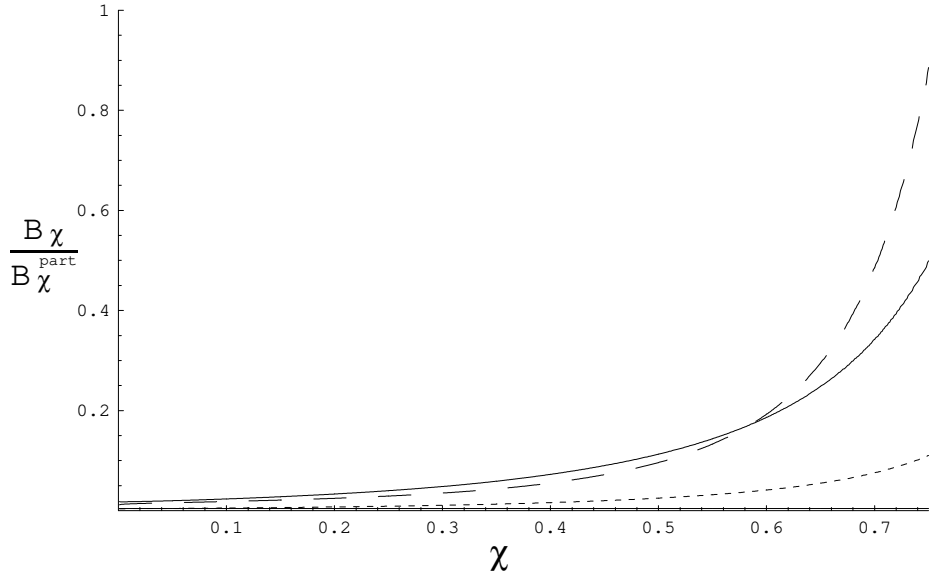}}
\caption{The fractional contributions with respect to
  the parton model result from the higher dimensional operators. The
  solid, dashed and dotted lines correspond to the contributions from
  $\lambda_2$, $\rho_1$ and $\rho_2$, respectively. The contribution
 from $\lambda_1$ is too small to be seen.}
\label{cut_plot} 
\end{figure}

The long distance $c \bar{c}$ resonances in the $d{\cal B}/d\hat{q}^2$
spectrum
must be cut out before the theory can be compared 
to measurements.
Thus we investigate 
the importance
of the nonperturbative corrections when we integrate only a fraction
of phase space $\hat{q}^2 > \chi$.  We define a partially integrated
branching ratio
\begin{equation}
{\cal B}_\chi = \frac{1}{{\cal B}_0} \int_\chi^1 \, d\hat{q}^2 
  \frac{d {\cal B}}{d\hat{q}^2} 
\end{equation} 
that depends on the size of the accessible phase space.  
Figure \ref{cut_plot} shows the fractional correction to the integrated parton 
level rate from each of the nonperturbative parameters $\lambda_i,\rho_i$
as a function of the minimum accessible dilepton invariant mass $\chi$.
In this plot we have
chosen the same values for the nonperturbative parameters as in Figure
\ref{diff_spec}. Over the entire range of $\hat{q}^2$ the contributions from ${\cal
  O}(1/m_b^3)$ operators are
comparable to the ones from ${\cal O}(1/m_b^2)$ operators. This
indicates that this decay is
unsuitable for extracting the matrix element $\lambda_1$ as has been
suggested in \cite{hiller_hqet}. This issue will be investigated more
in \cite{bigone}. Of course, the sizes of the $\rho_i$ 
contributions shown here should not be taken as accurate indications
of the actual size of the corrections, but rather as estimates of the
uncertainty in the prediction. We see that for $\chi \sim 0.75$ the
contribution from the $\rho_1$ matrix element is potentially of the same size as
the parton model prediction. This is a clear signal that the OPE is no 
longer valid. Even at $\chi \sim 0.5$ the
contribution is about 10\%, which is very interesting in light of the
fact that the CLEO search strategy
for this decay imposes the cut \cite{CLEO_search}
$\hat{q}^2 \ge \chi = (m_{\psi'} + 0.1\; \rm{ GeV})^2 / m_b^2 = 14.33 \, \rm{GeV}^2/m_b^2 = 0.59$, where we have used $m_b = 4.9 \, \rm{GeV}$.
Investigating this particular value of the cut
in more detail we find the individual contributions to the integrated
spectrum to be 
\begin{eqnarray}
  {\cal B}_{0.59} &=& 3.8 +
  1.9 \left(\frac{\lambda_1}{m_b^2} + \frac{{\cal T}_1 + 3 {\cal
        T}_2}{m_b^3}\right) - 134.7 \left( \frac{\lambda_2}{m_b^2} +
    \frac{{\cal T}_1 + 3 {\cal
        T}_2}{3 m_b^3}\right) \nonumber\\
&& + 614.9 \frac{\rho_1}{m_b^3} 
        + 134.7 \frac{\rho_2}{m_b^3} + 560.8 \frac{f_1}{m_b^3}.
\end{eqnarray}
To estimate the uncertainty
induced by the ${\cal O}(1/m_b^3)$ parameters we fix $\lambda_i$ to the values given above,
then randomly vary the magnitudes of the parameters
$\rho_i, {\cal T}_i$ and $f_1$ between $-(0.5 \, \rm{GeV})^3$ and $(0.5 \,\rm{GeV})^3$
as suggested by dimensional analysis.  We also impose positivity
of $\rho_1$ as indicated by the vacuum saturation approximation 
\cite{vac_sat}, and the constraint \cite{gremm}
\begin{equation}
\rho_2 - {\cal T}_2 - {\cal T}_4 = 
  \left( \frac{\alpha_s(m_c)}{\alpha_s(m_b)} \right)^{3/\beta_0} 
  \frac{ M_B^2 \Delta M_B (M_D + \bar{\Lambda}) - 
         M_D^2 \Delta M_D (M_B + \bar{\Lambda})}{M_B + \bar{\Lambda} - 
   \left( \frac{\alpha_s(m_c)}{\alpha_s(m_b)} \right)^{3/\beta_0}
   (M_D + \bar{\Lambda})}
\end{equation}
derived from the ground state meson mass splittings 
$\Delta M_H = M_{H^*} - M_H \;\; (H = B,D)$.  Here $\beta_0$ is the
well known coefficient of the beta function $\beta_0 = 11-\frac{2}{3} n_f$. Taking the 1 $\sigma$
deviation as a reasonable estimate of the uncertainties from ${\cal O}(1/m_b^3)$
contributions to the total rate at this cut, we again find the
uncertainty to be at the 10\% level. Relaxing the positivity constraint on $\rho_1$ enlarges the
uncertainty to about 20\%.

We have calculated the ${\cal O}(1/m_b^3)$ contributions to the
differential spectrum $d{\cal B}/d\hat{q}^2$. We have found that the
contributions of the new operators are small compared to the parton
level spectrum except close to the
endpoint $\hat{q}^2 \to 1$, where it is well known that the
convergence of the OPE breaks down. Due to large numerical
coefficients, however, the contributions from dimension six operators
are comparable to the contributions from the dimension five operators. We have also investigated the
uncertainties from the new nonperturbative operators on the total
decay rate evaluated with a lower cut $\chi$ on the dilepton invariant
mass. For $\chi \sim 0.59$, as proposed by CLEO, the uncertainties are 
around 10\%. Increasing the value of $\chi$ rapidly increases the
uncertainties on the partially integrated rate. At $\chi \sim 0.75$
this uncertainty is 100\%, signalling that the convergence of the OPE
has broken down.

We would like to thank Michael Luke for many discussions related to
this project.  This research was supported by the Natural Sciences
and Engineering Research Council of Canada.

\end{document}